\documentstyle[prl,aps,psfig]{revtex}

\begin{document}
\draft

\twocolumn[\hsize\textwidth\columnwidth\hsize\csname
@twocolumnfalse\endcsname

\widetext
\title{The Mott Metal-Insulator transition in the half-filled 
Hubbard model on the Triangular Lattice.}
\author{Massimo Capone, Luca Capriotti, and Federico Becca}
\address{International School for Advanced Studies (SISSA), 
and Istituto Nazionale per la Fisica della Materia (INFM) \\ 
Unit\`a Trieste-SISSA, Via Beirut 2-4, I-34014 Trieste, Italy}
\author{Sergio Caprara}  
\address{ Dipartimento di Fisica, Universit\`a di Roma ``La Sapienza'',
and  Istituto Nazionale per la Fisica della Materia (INFM) \\
Unit\`a  Roma 1, Piazzale Aldo Moro 2, I-00185 Roma, Italy}
\date{\today}
\maketitle

\begin{abstract}
We investigate the metal-insulator transition in the half-filled Hubbard model 
on a two-dimensional triangular lattice 
using both the Kotliar-Ruckenstein slave-boson 
technique, and exact numerical diagonalization of finite clusters. 
Contrary to the case of the square lattice, where the 
perfect nesting of the Fermi surface leads to a metal-insulator transition 
at arbitrarily small values of $U$, always accompanied by 
antiferromagnetic ordering, on the triangular lattice, due 
to the lack of perfect nesting, the transition takes place at a finite value 
of $U$, and frustration induces a non-trivial competition among different 
magnetic phases. 
Indeed, within the mean-field approximation in the slave-boson approach, 
as the interaction grows the paramagnetic metal turns into a metallic 
phase with incommensurate spiral ordering. Increasing further the interaction, 
a linear spin-density-wave is stabilized, and finally for strong coupling  the 
latter phase undergoes a first-order transition towards an antiferromagnetic 
insulator. 
No trace of the intermediate phases is instead seen in the 
exact diagonalization results, indicating a transition between a paramagnetic 
metal and an antiferromagnetic insulator.

\end{abstract}

]

\narrowtext

The Mott metal-insulator transition (MIT), i.e. the transition from a 
metallic  to an insulating phase driven by the electronic 
correlation \cite{mottmit,BR},
is one of the most relevant issues in condensed matter theory. 
In the last few years it has been also the object of an intensive study, 
due to many experimental evidences of Mott insulators ranging from the 
parent compounds of the superconducting cuprates \cite{parent} 
to the alkali fullerides of the type ${\rm A}_4 {\rm C}_{60}$ \cite{K4C60}.

The simplest model in which the competition between the delocalizing effect 
of the kinetic energy and the electronic correlation can give rise to a MIT 
is the Hubbard model
\begin{eqnarray}
{\cal H} = &-&t \sum_{<ij>\atop\sigma} 
\left(c_{i,\sigma}^{\dagger} c_{j,\sigma} +h.c.\right) 
\nonumber\\
&-&\mu\sum_{i,\sigma} c_{i,\sigma}^{\dagger} c_{i,\sigma}
+U\sum_{i}n_{i\uparrow}n_{i\downarrow},
\end{eqnarray} 
where $c_{i,\sigma}^{\dagger}$ ($c_{i,\sigma}$) creates (destroys) 
an electron with spin $\sigma$ on the site $i$ and $n_{i,\sigma} = 
c_{i,\sigma}^{\dagger}c_{i,\sigma}$ is the number operator; 
$t$ is the hopping amplitude, $U$ is the Hubbard on-site repulsion, 
$\mu$ is the chemical potential. 
The hopping is restricted to nearest-neighbors and the indices 
$i,j$ label  the points ${\bf r}_{i}$ and ${\bf r}_{j}$ of a d-dimensional 
lattice. 

At half-filling (i.e. for a number of electrons equal to the number of sites),
this model is known to undergo a MIT by increasing the interaction strength $U$. 
On the d-dimensional cubic lattice the perfect 
nesting property of the Fermi surface makes the model unstable towards 
antiferromagnetism as soon as a non-zero $U$ is turned on, driving the system 
to the insulating state. 
In this paper we focus on the triangular lattice as a prototype
for a model where the perfect nesting is absent for the uncorrelated
metal \cite{tosatti}. 
Since in the $U/t \to \infty$ (Heisenberg) limit the model
is likely to display a N\'eel ordered (insulating) ground state (GS) 
\cite{bernu,caprio}, 
the  MIT is expected to occur for finite $U$. 

Besides its theoretical relevance, our analysis has also an
experimental counterpart. In fact, 
the adlayer structures on semiconductor 
surfaces, such as ${\rm SiC}(0001)$ \cite{sic} or
${\rm K}/{\rm Si}(111):{\rm B}$ \cite{Ksi111},  
have recently turned out to be an almost ideal environment for the study of 
Mott insulators \cite{hellberg&C} and are
characterized by a $\sqrt{3}\times\sqrt{3}$ arrangement of the 
dangling-bond surface orbitals, which 
are likely to be well described 
by bidimensional strongly correlated Hamiltonians \cite{hellberg&C} 
on the triangular lattice.

Hartree-Fock (HF) calculations, performed by Krishnamurthy and co-workers 
\cite{KKK,KKK2}, 
produce a rather rich phase diagram: for small $U$ the system is a 
paramagnetic metal (PM),  which turns to a metal with 
incommensurate spiral spin-density-wave (Spiral Metal, SM) 
at $U = U_{c1} = 3.97t$.
Two successive first order transitions occur further increasing the
coupling: at $U = U_{c2} = 4.45t$ a semi-metallic linear spin-density wave 
(LSDW)  is stabilized, and a first order MIT to an 
antiferromagnetic insulator  (AFMI) occurs at  $U = U_{c3} = 5.27t$. 
In the same work it has also been argued that at finite 
temperature the model should present a Mott transition between a paramagnetic 
metal and a paramagnetic insulator.

However, the above transitions only occur at relatively large $U/t$ and the HF 
approximation is unreliable in the intermediate- and 
strong-coupling regime. 
Therefore we adopt the more appropriate slave-boson (SB) approach 
\cite{SB,KR} as an interpolating scheme between the $U/t=0$ and the 
$U/t\to\infty$ regimes. 
To allow for the presence of incommensurate spiral spin ordering, we 
introduce the spin-rotational invariant formulation \cite{AS} of the 
Kotliar-Ruckenstein SB approach \cite{KR}. 
The reader can find further details in Ref.~\cite{AS}.

We introduce on each site a set of four SB operators 
$e_i,s_{i,\uparrow},s_{i,\downarrow}$ and $d_i$ to label empty ($e$), singly 
($s$), and doubly ($d$) occupied sites, respectively. The spin projection 
$\varsigma=\uparrow,\downarrow$ is measured with respect to a local 
quantization axis, which is allowed to vary from site to site. The resulting 
SB Hamiltonian is
\begin{eqnarray}
{\cal H}=&-&t\sum_{<ij>\atop \varsigma,\varsigma'}\left[
{\tilde c}^\dagger_{i,\varsigma} z^\dagger_{i,\varsigma}
({\cal R}_i^\dagger {\cal R}_j)_{\varsigma,\varsigma'} 
z_{j,\varsigma'} {\tilde c}_{j,\varsigma'} +H.c.\right]
\nonumber\\
&-&\mu\sum_{i,\varsigma}
{\tilde c}^\dagger_{i,\varsigma} {\tilde c}_{i,\varsigma}
+U\sum_i d^\dagger_i d_i
\nonumber\\
&+&\sum_i \lambda_i \left( e^\dagger_i e_i+d^\dagger_i d_i +
\sum_\varsigma s^\dagger_{i,\varsigma} s_{i,\varsigma} -1 \right)
\nonumber\\
&+&\sum_{i,\varsigma}\Lambda_{i,\varsigma}\left(
{\tilde c}^\dagger_{i,\varsigma} {\tilde c}_{i,\varsigma} - 
s^\dagger_{i,\varsigma} s_{i,\varsigma} - d^\dagger_i d_i\right)
\label{hamilt}
\end{eqnarray}
where ${\tilde c}_{i,\varsigma},{\tilde c}^\dagger_{i,\varsigma}$ are the 
pseudofermion operators, the Lagrange multipliers $\lambda_i$ and 
$\Lambda_{i,\varsigma}$ enforce on each site the completeness constraint and 
the correct fermion counting respectively, the operator ${\cal R}_i$ 
rotates the local reference frame back to the laboratory frame, and the 
operator
$$
z_{i,\varsigma}={e^\dagger_i s_{i,\varsigma} + 
s^\dagger_{i,-\varsigma}d_i\over 
\sqrt{1-d^\dagger_i d_i -s^\dagger_{i,\varsigma}s_{i,\varsigma}}~
\sqrt{1-e^\dagger_i e_i -s^\dagger_{i,-\varsigma} s_{i,-\varsigma}}}
$$
reconstructs the hopping amplitude in the enlarged Fock space, and yields
the correct $U\to 0$ limiting behavior in the mean-field approximation 
\cite{KR,AS}. When the angle between two local quantization axes depends
only on their relative position, up to a global phase factor one can assume
${\cal R}_i^\dagger {\cal R}_j=\exp[ {\rm i}{\bf Q}\cdot
({\bf R}_i-{\bf R}_j) \tau_y/2]$, where $\tau_y$ is the Pauli matrix and 
${\bf Q}$ is the (incommensurate) modulating wavevector \cite{AS}. In such a 
case a mean-field description with real site-independent expectation values 
for the SB operators 
\begin{equation}
\langle e_i^{(\dagger)}\rangle=e_0;~~
\langle s_{i,\varsigma}^{(\dagger)}\rangle=s_{0,\varsigma};~~
\langle d_i^{(\dagger)}\rangle=d_0,
\label{sbmf}
\end{equation}
and for the Lagrange multipliers
\begin{equation}
\langle \lambda_i\rangle=\lambda_0;~~
\langle \Lambda_{i,\varsigma}\rangle=\Lambda_{0,\varsigma},
\label{lmmf}
\end{equation}
is possible.
Eqs. (\ref{sbmf}) and (\ref{lmmf}) refer to the case in which 
the translational symmetry is not broken and the expectation 
values of the bosons and of the Lagrange multipliers do not depend
on the site.
We have also studied configurations with broken translational symmetry.
In particular we considered solutions in which in the bosons have 
different values on each of the three sublattices, and 
analogous to LSDW found in Hartree-Fock\cite{KKK2}. The latter solutions 
can be found considering a four-sites unit cell.
A similar SB calculation has been performed in Ref. \cite{trumper},
where, however, the generalization of the SM phase found in HF 
was never recovered as an energy minimum.
In the case of spiral spin ordering,
the Hamiltonian (\ref{hamilt}) can be analytically diagonalized 
by adopting the Bloch 
representation, and performing a unitary transformation with respect to spin 
indices, yielding
\begin{eqnarray*}
&~&E_{\bf k,\pm}={1\over 2}\left[t(z_{0,\uparrow}^2+z_{0,\downarrow}^2)T_e
+\Lambda_{0,\uparrow}+\Lambda_{0,\downarrow}\right]-\mu \nonumber\\
&\pm& {1\over 2}
\sqrt{\left[t(z_{0,\uparrow}^2-z_{0,\downarrow}^2)T_e
+\Lambda_{0,\uparrow}-\Lambda_{0,\downarrow}\right]^2+4t^2 z_{0,\uparrow}^2
z_{0,\downarrow}^2 T_o^2}
\nonumber
\end{eqnarray*}
where 
$T_e=-\sum_{\bf l}\cos({\bf Q}\cdot {\bf l}/2)\cos({\bf k}\cdot{\bf l})$, 
$T_o=-\sum_{\bf l}\sin({\bf Q}\cdot {\bf l}/2)\sin({\bf k}\cdot{\bf l})$,
and ${\bf l}=(1,0),(1/2,\pm\sqrt{3}/2)$ are the nearest-neighbor displacements. 
The self-consistency equations are obtained by minimizing the free energy
$$
{\cal F}={\cal F}_0-T\sum_{{\bf k},\alpha=\pm} \log \left(
1+{\rm e}^{-E_{{\bf k},\alpha}/T}\right),
$$
where ${\cal F}_0=N[Ud_0^2+\lambda_0(e_0^2+d_0^2+s_{0,\uparrow}^2
+s_{0,\downarrow}^2-1)-\Lambda_{0,\uparrow}(d_0^2+s_{0,\uparrow}^2)
-\Lambda_{0,\downarrow}(d_0^2+s_{0,\downarrow}^2)+\mu n]$, $N$ is the
number of sites, and $n$
is the electron density per site, and read
\begin{equation}
\label{sceq}
{\partial {\cal F}_0\over \partial {\cal X}}+\sum_{{\bf k},\alpha=\pm}
{\partial E_{{\bf k},\alpha}\over \partial {\cal X}} f(E_{{\bf k},\alpha})=0,
\end{equation}
where $f(E)=[{\rm e}^{E/T}+1]^{-1}$ is the Fermi function and ${\cal X}$
represents generically one of the parameters (\ref{sbmf}), (\ref{lmmf}) and 
the two components of the pitch vector ${\bf Q}$. The chemical potential 
$\mu$ is fixed by the condition
$$
\sum_{{\bf k},\alpha=\pm}f(E_{{\bf k},\alpha})=nN.
$$
In this paper we assume henceforth $n=1$ (half-filling).

\begin{figure}
\centerline{\psfig{bbllx=80pt,bblly=200pt,bburx=510pt,bbury=575pt,%
figure=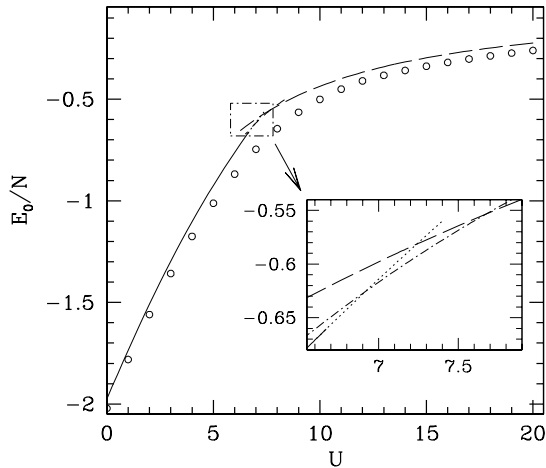,width=70mm,angle=0}}
\caption{$U$ dependence of the ground-state energy per site. SB results:
PM (solid line), SM (dotted line), LSDW (dot-dashed line), and AFMI (dashed line). Open dots are the
exact diagonalization results for the $N=12$ cluster.
\label{fig_ene}
}
\end{figure}

The self-consistency equations (\ref{sceq}) yield the same solutions found in 
HF, namely a paramagnetic metal, a metal with incommensurate spiral ordering, 
a linear spin-density-wave and an antiferromagnetic insulator. 
As in HF, the PM-SM transition is 
continuous, and the other two transitions are of first order, 
but all of them  occur at larger coupling values, 
$U_{c1} = 6.68t$, $U_{c2}= 6.84t$,  and $U_{c3} = 7.68t$.  The energy 
curves corresponding to the above phases are reported in Fig.~\ref{fig_ene}.
Our results agree with Ref. \cite{trumper} as far as the PM, AFMI, 
and LSDW phases
are concerned, but we also find a region of stability for the SM phase,
which was not detected in Ref. \cite{trumper}.
These authors  were indeed looking for spiral phases starting from
the strong-coupling side, and following them to weaker coupling.
On the other hand, our analysis shows that a spiral phase develops
continuously from the PM at intermediate coupling and it ends in
a critical point soon after the level-crossing with the AFMI (see the inset
in Fig.~\ref{fig_ene}), and does not exist at strong-coupling.
Therefore, our SM phase is the generalization of the corresponding
phase found within HF \cite{KKK2}, 
and it is unrelated to the high-energy SM phases
of Ref. \cite{trumper}.
However, the region of existence of the SM
is narrower within SB as compared to HF, and the magnetization 
$m = {1\over 2}(n_{\uparrow} - 
n_{\downarrow})$ is always less than 0.1, a really small value with respect to 
the HF value ($\le 0.4$). Therefore the jump of the magnetization at the 
SM-AFMI transition is substantially larger than in the HF approximation.
We point out that, contrary to nesting models, where the presence of free 
particles (doping) is a necessary condition for spiral ordering 
\cite{AS,TUG,CAT}, here the spiral phase exists at half-filling, as previously 
shown in Ref. \cite{KKK}, within the HF approximation. 
Despite the overall qualitative agreement between the HF and the SB phase
diagrams, the main outcome of the comparison between them is that the
stability of the SM phase
is strongly reduced. Furthermore, the SM is hardly 
distinguishable from the PM  in its whole region of stability. 
It is reasonable to expect that the inclusion
of quantum fluctuations washes out these phases leaving the way open for 
a transition between a PM and the AFMI.

Despite the strong frustration of the antiferromagnetic (AFM) order 
on the triangular 
lattice \cite{bernu,caprio}, both the HF and SB approaches indicate 
no paramagnetic Mott insulating phase                                         
in the zero-temperature phase diagram of the half-filled Hubbard model.
      
In particular, within the SB approach, we can indicate how far the system is 
from the Brinkman-Rice transition \cite{BR} to a paramagnetic Mott insulator. 
In fact, if the possibility for magnetic ordering is neglected, 
the paramagnetic metallic 
phase undergoes a Brinkman-Rice transition with vanishing double occupancy 
and effective hopping amplitude, for a critical value (at $T=0$) of the Hubbard 
interaction $U_{BR}=32tN^{-1}\sum_{\bf k} \varepsilon_{\bf k} 
\Theta(2t\varepsilon_{\bf k}+\mu)$ ($\epsilon_k = -T_e({\bf Q}=0)$) 
\cite{BR,KR}, 
i.e. $U_{BR}/t\simeq 15.8$ on the triangular lattice.
As we see this value is much larger than $U_{c1}/t$, $U_{c2}/t$ and
$U_{c3}/t$ found  above. The system is therefore not even close 
to the  Brinkman-Rice transition when the MIT occurs.

\begin{figure}
\centerline{\psfig{bbllx=80pt,bblly=250pt,bburx=510pt,bbury=575pt,%
figure=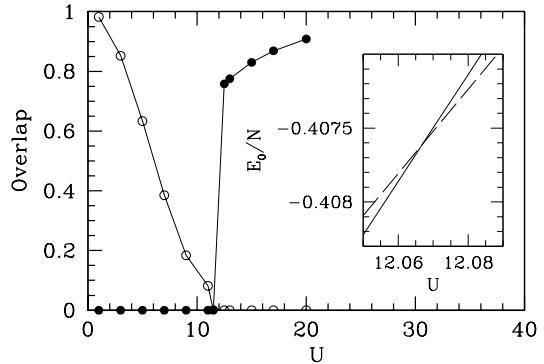,width=70mm,angle=0}}
\caption{Overlap of the finite-$U$ GS with the $U=0$
(empty dots) and the $U=100t$ (full dots) GS's, for $N=12$.
In the inset the GS energy per site in the
${\bf k}=(0,0)$ (solid line)  and ${\bf k}=(2\pi/3,0)$
(dashed line) subspaces is plotted versus $U$.
\label{fig_over}
}
\end{figure}   

In order to understand to which extent  the picture we found within the 
mean-field SB theory survives in an exact treatment of the model, we performed
exact diagonalization  of small clusters by means of the standard Lanczos 
algorithm. The largest lattice compatible with all the symmetries of the model 
that can be handled with exact diagonalization is a $N=12$ site 
cluster \cite{bernu}. We always used 
twisted boundary conditions  with a suitable phase such that the 
half-filled system is in a closed-shell configuration. 
This is important in order to perform a reasonable investigation
of the conduction properties of the finite-size system.
It turns out that the boundary conditions that minimize the energy in a closed-shell configuration 
for $U = 0$ leave the system in a closed-shell configuration at all $U$. 
The energy is shown as a function of $U$ in Fig.~\ref{fig_ene}. The overall 
agreement with the mean-field SB results is good, the largest deviations 
($\sim 20\%$) being, as expected, at intermediate coupling ($U/t \sim 7$).

To check the occurrence of a discontinuous phase transition 
we evaluated the overlap between the GS wave function and the 
two limiting cases of $U=0$, and for large $U$ (namely, $U=100t$). 
As shown in Fig.~\ref{fig_over} 
on the large-$U$ side of the diagram the GS has a large overlap 
to the AFM strong-coupling state and a 
vanishing overlap with 
the non-interacting metallic one. 
On the metallic side the overlap with the non-interacting state is 
always finite, but it is a decreasing function of $U$; in this regime the GS 
has anyway a vanishing overlap to the AFM state. We have therefore a clear 
evidence for  a strongly correlated metal with a decreasing coherent part. 
In particular the sharp change of the GS wave function at $U_{MIT} \simeq 12.07t$ is 
due to a level-crossing occurring between a metallic and 
an antiferromagnetic solutions, as it is shown in the inset
of Fig.~\ref{fig_over}.
These results, however, do not rule out the possibility of a continuous 
transition within the metallic phase, i.e. 
the PM-SM transition found with SB.

\begin{figure}
\centerline{\psfig{bbllx=80pt,bblly=225pt,bburx=510pt,bbury=575pt,%
figure=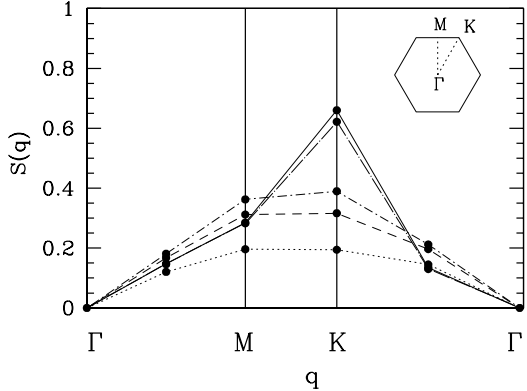,width=70mm,angle=0}}
\caption{Spin structure factor $S({\bf q})$: U=2t (dotted line), U=8t (dashed line),
U=11.5t (dot-short-dashed line), U=12.5t (dot-long-dashed), U=40t (solid line).
\label{fig_sq}
}
\end{figure}  

In Fig.~\ref{fig_sq} we show the spin structure factor, 
$S({\bf q})=\sum_{i,j}S^z_iS^z_j \exp[{{\bf q}\cdot({\bf r}_i-{\bf r}_j) }]/N$, 
for different values of $U$.
The results do not suggest any intermediate state between a metallic state 
without magnetic order and the AFM insulator, as $S({\bf q})$ abruptly
changes from a structureless behavior to an AFM pattern peaked at the
classical ordering wavevector, i.e.  ${\bf Q_0}=(4\pi/3,0)$.
Although we suspect that the intermediate phases are an artifact of the
mean-field approach, the weakness and the strong size-dependence 
of the spiral phases suggested by the SB results, may make them
unaccessible on our 12-site lattice.

\begin{figure}
\centerline{\psfig{bbllx=80pt,bblly=235pt,bburx=510pt,bbury=575pt,%
figure=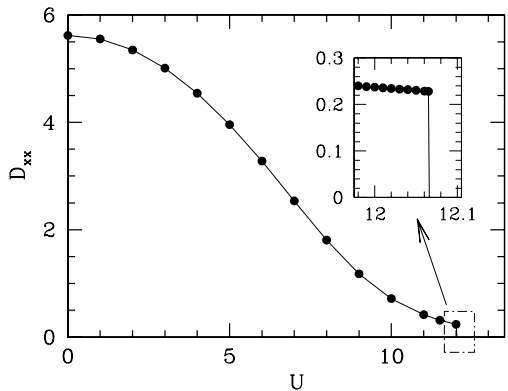,width=70mm,angle=0}}
\caption{Exact calculation of the Drude weight as a function of $U$
for the $N=12$ cluster.
\label{fig_drude}
}
\end{figure}

Using the Lanczos algorithm we have also calculated the finite-frequency 
optical conductivity $\sigma(\omega)$ and the Drude weight, measuring the 
electronic mobility.
The real part of the $xx$ component of the conductivity tensor for 
a tight-binding model at zero 
temperature may be expressed in terms of the Kubo formula \cite{maldague}
\begin{equation}
\sigma_{xx}(\omega) = D_{xx}\delta(\omega) + 
\Im \langle 0\vert J_{x}^\dagger 
{1\over \omega-{\cal H}+E_0-i\delta}J_{x}\vert 0\rangle
\label{kubo},
\end{equation}
where $J_x= \sum_{i,\sigma,\bf{l}} l_x (c^{\dagger}_{i,\sigma} 
c_{i+{\bf{l}},\sigma} - h.c.)$ is the $x$-component of the current operator. 
The coefficient of the zero-frequency delta function contribution $D_{xx}$,  
the Drude weight, is given by the f-sumrule\cite{maldague} 
\begin{equation}
\label{drude}
D_{xx} = -{\pi e^2\over 2}\langle {\cal H}^t_x \rangle - 
\sum_{n\neq 0}{\vert\langle\phi_0
\vert J_x\vert\phi_n\rangle\vert^2\over E_n - E_0},
\end{equation}
where ${\cal H}^t_x = \sum_{i,\sigma,\bf{l}} l_x^2 (c^{\dagger}_{i,\sigma} 
c_{i+{\bf{l}},\sigma} + h.c.)$, and $\vert \phi_n\rangle$ is the 
eigenfunction of ${\cal H}$ with eigenvalue $E_n$.

The latter quantity, which is reported in Fig.~\ref{fig_drude} is a  
direct measure of the metallic character of the state, and the MIT
is signaled by the vanishing of $D_{xx}$ \cite{kohn}. 
For a finite system, $D_{xx}$ does not vanish 
for any value of $U$, but an abrupt change takes place at the 
level-crossing point. 
For $U < U_{MIT}$, $D_{xx}$ is a decreasing function
of the interaction, which resembles the overlap in Fig. \ref{fig_over}.
An abrupt change takes place at the
level-crossing point and for $U > U_{MIT}$ it becomes negative, 
a common phenomenon in the insulating phase of 
a small-size system \cite{dagotto}.

All the results of exact diagonalization point towards the same
direction: the metal-AFMI level-crossing found within the
SB mean-field approach is shifted to larger values of $U$.
The metallic solution exhibits a continuous loss of metallicity
with increasing $U$. 
The Drude weight is finite up to the MIT on the 12-site lattice although
it is quite small ($4\%$ of the non-interacting value).
We remark that, due to finite-size effects, we cannot exclude the
possibility that $D_{xx}$ vanishes before the transition to the AFMI is
reached. In such a case, there would be a region of parameters
in which the paramagnetic insulator exist, though the SB results
point in the opposite direction.

In conclusion, using the slave boson technique and the exact diagonalization, 
we have investigated the zero-temperature phase diagram
of the half-filled Hubbard model on a two dimensional triangular lattice.
The mean-field SB approach displays a rich phase diagram which
qualitatively resembles the one from HF calculations, but, on the other
hand, drastically reduces the stability of the spiral metal and
of the linear spin-density-wave states.
Namely, the weak-coupling paramagnetic metal continuously evolves into
a spiral metal at $U = U_{c1} = 6.68t$, which crosses the linearly
polarized spin-density-wave ground-state at $U = U_{c2} = 6.84t$.
The latter phase undergoes a further first-order transition
towards an antiferromagnetic insulator at $U = U_{c3} = 7.68t$.
All these transitions occur for coupling constants
substantially smaller than the critical value for the Brinkman-Rice
transition to a paramagnetic insulator ($U_{BR} = 15.8t$).
The exact-diagonalization results present a first order transition 
between the paramagnetic metal and the antiferromagnetic insulator at $U_{MIT}=12.07 t$,
without intermediate ``exotic'' phases.

\begin{acknowledgments}

It is a pleasure to thank S. Sorella and A. Parola for suggestions and
fruitful discussions.  
Useful communications with A. E. Trumper are also acknowledged.
S.C. acknowledges the kind hospitality of the International School for 
Advanced Studies in Trieste, where most of this work was carried out.
This work was partially supported by MURST (COFIN99). 

\end{acknowledgments}

\end{document}